\documentclass[final,5p,times,twocolumn]{elsarticle}

\usepackage{amssymb}
\usepackage{mathtools}
\usepackage{multirow}

\journal{IEEE Sensors Journal}

\begin{document}

\begin{frontmatter}



\title{Understanding the Nonlinear Response of SiPMs}


\author{Víctor Moya}
\author{Jaime Rosado}

\affiliation{organization={IPARCOS-UCM, Instituto de Física de Partículas y del Cosmos and EMFTEL Department, Universidad Complutense de Madrid},
            city={Madrid},
            postcode={E-28040}, 
            country={Spain}}

\begin{abstract}
A systematic study of the nonlinear response of Silicon Photomultipliers (SiPMs) has been conducted through Monte Carlo simulations. SiPMs have been proven to show a universal nonlinear response when it is expressed in terms of relative parameters independent of both the gain and the photon detection efficiency (PDE). Nonlinearity has been shown to mainly depend on the balance between the photon rate and the pixel recovery time. However, exponential-like and finite light pulses have been found to lead to different nonlinear behaviors, which also depend on the correlated noise, the overvoltage dependence of the PDE, and the impedance of the readout circuit. Correlated noise has been shown to have a minor impact on nonlinearity, but it can significantly affect the shape of the SiPM output current.

Considering these dependencies and previous statistical analysis of the nonlinear response of SiPMs, two simple fitting models have been proposed for exponential-like and finite light pulses, explaining the role of their various terms and parameters. Even though these models have only three fitting parameters, they provide an accurate description of the nonlinear response of SiPMs for a wide range of situations.
\end{abstract}





\end{frontmatter}


\section{Introduction}
\label{sec:introduction}

Silicon photomultipliers (SiPMs) are solid-state photodetectors that offer excellent characteristics, including high gain, fast timing properties, very good photon-counting resolution, and high quantum efficiency \cite{renker2006geiger}. Additionally, they are compact, relatively affordable, operate at low voltage (a few tens of volts), and are insensitive to magnetic fields. Due to their numerous advantages over classical photomultiplier tubes (PMT), SiPMs are increasingly utilized in various fields, such as high-energy physics experiments \cite{CTA_SST, CTA_LST}, medical imaging \cite{gundacker2020silicon}, biophotonics \cite{bruschini2019single}, light detection and ranging (LiDAR) systems \cite{agishev2013lidar}, and more.

A SiPM consists of an array of Geiger-mode avalanche photodiodes (G-APDs), hereafter called pixels, connected in parallel to a common readout, so that the SiPM output signal is the sum of the signals from all the pixels. The device is biased above the breakdown voltage, allowing a single incident photon to trigger a breakdown avalanche of high multiplication, producing a measurable signal. In ideal conditions, when a light pulse illuminates a SiPM, the output charge should be directly proportional to the number of detected photons. However, SiPMs exhibit a nonlinear response. This is because SiPMs have a finite number of pixels, and each pixel takes some time to recover after a breakdown avalanche. A fraction of photons can interact with unrecovered pixels, which have lower trigger probability and gain, resulting in a SiPM signal with lower amplitude than expected. Nonlinearity depends on the photon rate (i.e., both the amplitude and width of the light pulse) and the recovery time of pixels in a complex way. Additionally, SiPMs exhibit both correlated and uncorrelated noise, further complicating their performance \cite{rosado2015characterization, gallego2013modeling, vinogradov2012analytical}.

An exact analytical treatment of the nonlinear response of SiPMs is not possible. Nevertheless, several statistical models relying on simplifications of different aspects of the SiPM response are available \cite{rosado2019modeling, jeans2016modeling, vinogradov2016nonlinearity, vinogradov2015nonlinearity, van2010comprehensive}. In \cite{rosado2019modeling}, a comprehensive statistical model of the SiPM response for light pulses of arbitrary shape and duration was presented. This model provides a simple expression for the mean output charge of a SiPM, which depends on only two parameters that can be measured or fitted. However, due to the adopted approximations, the applicability of this model is limited to situations where nonlinear effects and correlated noise are moderate. Besides, the model cannot describe the statistical fluctuations of the output charge or nonlinear effects on the signal shape.

On the other hand, a Monte Carlo (MC) treatment can potentially include all factors that affect SiPM performance, allowing a detailed simulation of the SiPM response for arbitrary light pulses. Various MC codes are available on different platforms \cite{mehadji2022simulating, SimSiPM, jha2013simulating}. These codes mainly differ in the electrical model of the SiPM for signal formation and pixel recovery, in the modeling of correlated noise, and in the description of both the trigger probability and gain of recovering pixels. The more details included in the simulation, the more input parameters are required, some of them not known with sufficient precision. Consequently, even though a very detailed MC simulation is used, parameter tuning or fitting to experimental results may be necessary to reproduce the actual response of a SiPM in a practical situation.

In this work, a hybrid strategy was followed. Firstly, we used MC simulations to conduct a systematic analysis of the different factors affecting the nonlinear response of SiPMs, regarding both the output charge and the signal shape. To this end, we utilized the Matlab MC code developed by Abhinav et al \cite{jha2013simulating}, which was specially designed to account for the nonlinear effects in SiPMs. However, we modified this code to simulate correlated noise more accurately and simplified most features of the electrical model implemented by the authors of the code to understand better the role of the main SiPM parameters in the nonlinear response. Secondly, based on our previous statistical model described in \cite{rosado2019modeling}, we found simple analytical expressions, depending on a few parameters, that fit the simulation results of the mean output charge for light pulses of different shapes and arbitrary intensity over a very wide range of SiPM parameters. The proposed models provide a simple but accurate description of the performance of SiPMs in the nonlinear region, showing clearly the relationships between the many variables of the problem.

\section{Overview of the SiPM Response}\label{sec:overview}

In this section, the SiPM performance is described with an emphasis on the effects of pixel recovery and correlated noise. The main features affecting the nonlinear response are identified and parameterized.

Let us assume a SiPM biased at a certain overvoltage, referred to as the operation overvoltage \(U_{\rm op}\). When the SiPM is in a steady state, i.e., no signal is produced, all the pixels are biased at the same overvoltage \(U_{\rm op}\). If a photon interacts with a pixel, it can trigger a self-sustaining charge avalanche in the diode PN junction. A current \(I_{\rm q}\) flows through the quenching resistor \(R_{\rm q}\) connected in series with the diode, causing the voltage at the diode to decrease to a value close to the breakdown voltage, and the avalanche stops. The current continues to flow until the diode overvoltage \(U\) is restored to \(U_{\rm op}\). The diode overvoltage as a function of the time elapsed from the photon arrival is therefore given by
\begin{equation} \label{overvoltage recovery}
    U(t)=U_{\rm op}-I_{\rm q}(t) \, R_{\rm q}\,.
\end{equation}

The avalanche duration is sub-nanosecond, while the recovery of the diode overvoltage typically takes a few tens of nanoseconds. Assuming that the avalanche is instantaneous and the diode overvoltage drops to \(0\) at that moment, the current at the quenching resistor is
\begin{equation} \label{I_q}
    I_{\rm q}(t)=\frac{U_{\rm op}}{R_{\rm q}}\,\exp\left(-\frac{t}{\tau_{\rm rec}} \right)\,.
\end{equation}
The time constant \(\tau_{\rm rec}\), called recovery time, is approximately given by \(R_{\rm q} C_{\rm d}\), where \(C_{\rm d}\) is the diode capacitance. The total charge released by the pixel is \(C_{\rm d} U_{\rm op}\), and therefore, the gain (i.e., the avalanche multiplication) is \(G=C_{\rm d} U_{\rm op}/e\), where \(e\) is the elementary charge. The gain is typically \(10^{5} - 10^6\) and is customarily characterized as a function of \(U_{\rm op}\) at single-photon conditions.

If a photon triggers an avalanche in a pixel with overvoltage \(U<U_{\rm op}\), i.e., while it is recovering from a previous avalanche, the diode overvoltage drops from \(U\) to \(0\), the current at the quenching resistor rises to \(U_{\rm op}/R_{\rm q}\) again, and the pixel recovery restarts. Therefore, the multiplication of the second avalanche is
\begin{equation}\label{gain_recovery}
    G'=C_{\rm d} U/e=G\,U/U_{\rm op}\,.
\end{equation}

\begin{figure}[h]
\centering
\includegraphics[width=\linewidth]{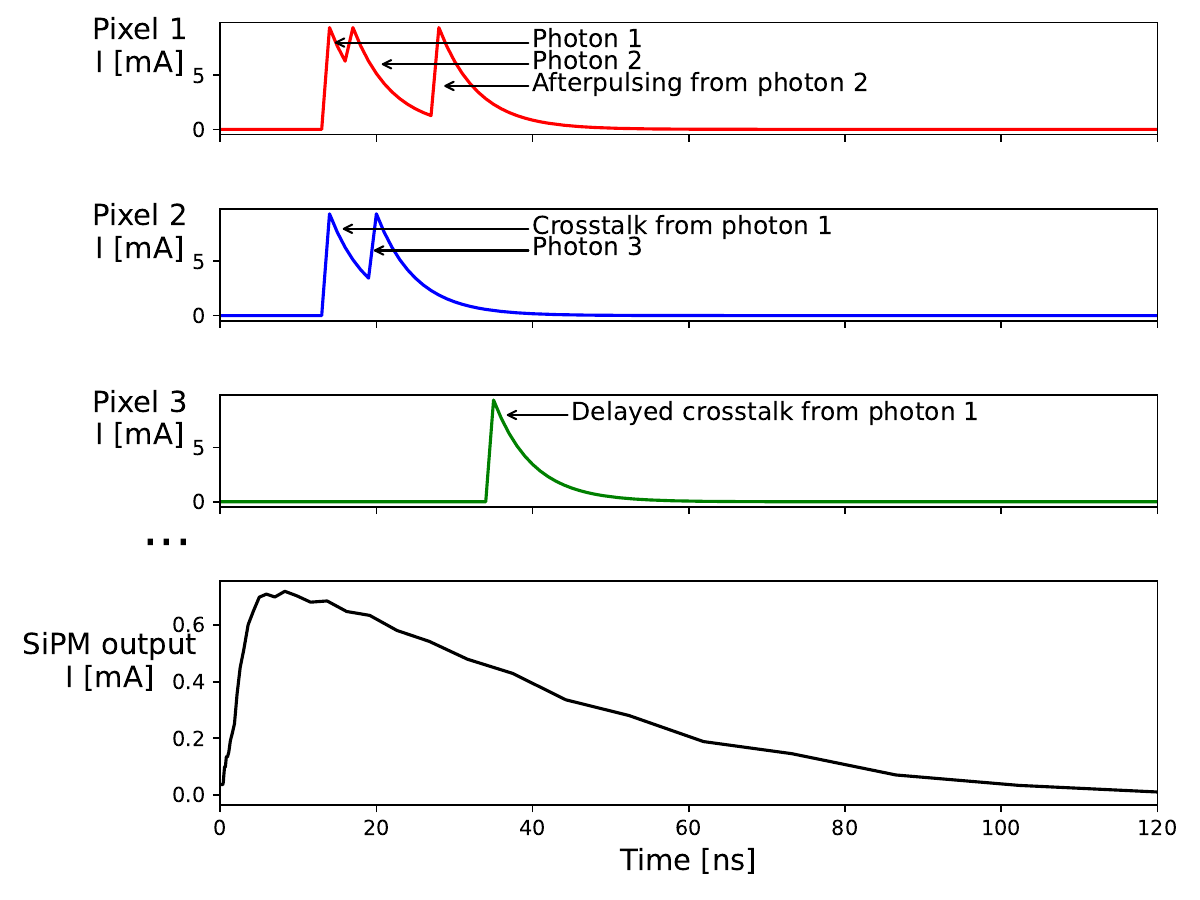}
\caption{Simulated current signals generated by three pixels of a SiPM and total output signal. The origin of each avalanche is identified in the simulation, although they are indistinguishable from each other. The effect of the pixel recovery in the avalanche amplitude can also be appreciated in the signals from pixels 1 and 2.} 
\label{fig:current_sum}
\end{figure}

The output current of a SiPM is the sum of the currents of all the individual pixels, as illustrated in Fig. \ref{fig:current_sum}. When the SiPM is illuminated with light pulses, the signal is usually integrated to measure the output charge \(Q\) for each pulse. For convenience, we define the parameter \(y=\langle Q\rangle/(N\,G\,e)\), where \(N\) is the number of pixels of the SiPM. This parameter \(y\) represents the equivalent number of full charge avalanches that should be produced by every pixel to add up the measured mean output charge \(\langle Q\rangle\). Notice that \(y\) is independent of gain.

A detailed electrical model of a SiPM, such as the one described in \cite{jha2013simulating}, should include the time evolution of the charge avalanche and the parasitic capacitance of the connections and the quenching resistor, which basically affect the shape of the output signal. In addition, the readout circuit has a small resistance that is usually neglected. However, if the output current is high enough, the voltage drop at this resistance may be significant, causing the overvoltage at all the pixels to decrease, and the multiplication of subsequent avalanches is reduced, contributing to nonlinearity.

The probability that an incident photon causes a charge avalanche is called photodetection efficiency (PDE), which is the product of the geometric efficiency, the quantum efficiency, and the trigger probability. The geometric efficiency is the probability that a photon incident in the sensitive SiPM surface arrives at a pixel, excluding the inactive regions between pixels. The quantum efficiency is the probability that a photon incident in a pixel creates an electron-hole pair inside the depleted volume around the PN junction. The trigger probability is the probability that free carriers generated in the depleted volume cause a successful charge avalanche. These factors, especially the trigger probability, depend on \(U_{\rm op}\). In this work, we assume the following expression for the PDE:
\begin{align} \label{PDE}
\varepsilon(U_{\rm op})&=\varepsilon_{\rm max}\,\left[1-\exp\left(-\frac{U_{\rm op}-U_0}{U_{\rm ch}}\right)\right]\;\;{\rm if}\; U_{\rm op}>U_0\nonumber\\
\varepsilon(U_{\rm op})&= 0\;\;{\rm if}\; U_{\rm op}\leq U_0\,,
\end{align}
which fits well with experimental data for many SiPMs (see, e.g., \cite{OTTE2017106}). In this equation, \(\varepsilon_{\rm max}\) is the saturation PDE value, \(U_{\rm ch}\) is a characteristic overvoltage value that determines how fast the PDE grows with \(U_{\rm op}\), and \(U_0\) is a small overvoltage threshold that has to be included to fit the PDE data for some SiPMs. Other expressions to describe the PDE are used in the literature, but the conclusions of this work are independent of the particular expression used. The PDE is also a function of the wavelength of incident light. However, we did not consider this dependence in our analysis.

As with the gain, the PDE is characterized at low light intensity conditions, at which all the pixels can be assumed to be in a steady state. However, the trigger probability in a recovering pixel with overvoltage \(U<U_{\rm op}\) is smaller. It can be assumed to grow with \(U\) at the same rate as the PDE grows with \(U_{\rm op}\) in (\ref{PDE}). For simplicity, this can be modeled in two steps. First, an incident photon can produce an avalanche ``seed" in a pixel with a probability equal to \(\varepsilon(U_{\rm op})\), regardless of whether the pixel is in a steady state or not. Second, if a seed is produced, it can trigger a charge avalanche with a probability \(\varepsilon(U)/\varepsilon(U_{\rm op})\).

The nonlinear response of a SiPM is conveniently described as a function of the mean number of seeds produced by photons in each pixel, which is given by the parameter \(x = \varepsilon(U_{\rm op})\langle n\rangle/N\), where \(\langle n\rangle\) is the mean number of photons incident in the SiPM per light pulse. Notice that \(y=x\) for an ideally linear response, but \(y<x\) when nonlinearity is shown. In the case of an instantaneous light pulse (i.e., no pixel recovery between two consecutive photons in a pixel), it is well-known that the SiPM response is simply given by
\begin{equation} \label{instantaneous}
y=1-\exp(-x)\,.
\end{equation}
However, for non-instantaneous pulses, the SiPM is much more complex, and \(y\) can be greater than unity.

In addition, SiPMs present both correlated and uncorrelated noise, which produce charge avalanches that are indistinguishable from the ones produced by photons, making the SiPM response even more complex. Uncorrelated noise is caused by thermally generated electron-hole pairs inside the pixel active region. In our approach, this can be modeled as a Poissonian generation of seeds uniformly distributed among all the pixels. However, the uncorrelated noise rate per pixel is typically below kHz, much smaller than \(1/\tau_{\rm rec}\).  Therefore, this effect has a negligible contribution to the nonlinear response of SiPMs.

On the other hand, correlated noise is the stochastic production of secondary avalanches induced by a primary avalanche. This effect scales with the signal amplitude and thus might be relevant to nonlinearity. There are three components of correlated noise (see Fig. \ref{fig:current_sum}). Prompt crosstalk is the nearly simultaneous production of secondary avalanches in neighboring pixels to the pixel where the primary avalanche took place, and it is due to the emission of infrared photons in this primary avalanche (see \cite{gallego2013modeling} and references therein). Delayed crosstalk is caused by minority carriers that diffuse in the silicon substrate until they reach a neighboring pixel, triggering secondary avalanches with some delay with respect to the primary avalanche \cite{rosado2015characterization}. Finally, afterpulsing is similar to delayed crosstalk, but the secondary avalanches are produced in the same pixel where the primary avalanche took place, usually while it is still recovering.

To model the correlated noise, a full charge avalanche is supposed to produce a random number of seeds of each noise component \(i\) (ct, dct, or aft) following a Poisson distribution with a mean of \( \lambda_i\). In the case that an avalanche is produced in a recovering pixel with overvoltage \(U\), the mean number of seeds of each noise component that it produces is \( \lambda'_i = \lambda_i \, U / U_{\rm op}\). For both crosstalk components, we assume that the generated seeds are uniformly distributed among the 4 closest neighbors of the primary pixel, which is an approximation that generally reproduces well the experimental data, as shown in \cite{gallego2013modeling, rosado2015characterization}. For prompt crosstalk, the small delay of the generated seeds with respect to the primary avalanche can be ignored. On the other hand, we assume the following time distribution for afterpulsing and delayed crosstalk \cite{rosado2015characterization}:
\begin{align} \label{afterpulsing time}
     \frac{{\rm d}\lambda_i}{{\rm d}t}(t) &= C_i \, t^{a_i} \, \exp\left(-\frac{t}{\tau_{\rm bulk}}\right)\;\;{\rm if}\; t>t_{\rm min}\nonumber\\
     \frac{{\rm d}\lambda_i}{{\rm d}t}(t) &= 0\;\;{\rm if}\; t\leq t_{\rm min}\,,
\end{align}
where \(\tau_{\rm bulk}\) is the minority carrier lifetime in the silicon bulk, which may range from a few ns to hundreds of ns, \(a_{\rm aft}=-1\) for afterpulsing, \(a_{\rm dct}=-0.5\) for delayed crosstalk, \(C_i\) is a normalization constant such that \(\lambda_i = \int_{t_{\rm min}}^{\infty} \frac{{\rm d}\lambda_i}{{\rm d}t}(t) \, {\rm d}t\), and \( t_{\rm min}\) is the minimum time delay between two avalanches that the readout circuit can distinguish, which is typically around \(5\)~ns. This model assumes that afterpulsing is caused by charge carriers diffusing in the silicon bulk, but afterpulsing can also be due to trapping and subsequent release of charge carriers in the depleted volume. Nevertheless, the particular shape of the delay time distribution is not relevant to describe nonlinear effects.

Each seed, regardless of whether it is generated by an incident photon or a noise event, is assumed to be able to trigger a charge avalanche with a probability \(\varepsilon(U)/\varepsilon(U_{\rm op})\), with \(U\) being the overvoltage of the pixel where the seed is generated. If an avalanche is triggered, it can produce further seeds via correlated noise in a cascade process.

Correlated noise is usually characterized at single-photon conditions, measuring the probability \(P_i\) that a primary avalanche produces one or more secondary avalanches of each type. For both crosstalk components, this probability is related to the above-mentioned parameter \(\lambda_i\) by
\begin{equation} \label{No crosstalk probabilit}
    P_i = 1 - \exp\left( -\lambda_i \right)\,.
\end{equation}
On the other hand, the trigger probability of an afterpulsing seed is a function of the delay time, because the pixel is recovering from the primary avalanche. The afterpulsing probability is related to \(\lambda_{\rm aft}\) by
 \begin{equation} \label{no afterpulsing probability}
    P_{\rm aft} = 1 - \exp\left[- \int_{t_{\rm min}}^{\infty} \frac{{\rm d}\lambda_{\rm aft}}{{\rm d}t}(t) \, \frac{\varepsilon(U(t))}{\varepsilon(U_{\rm op})} \, {\rm d}t  \right]\,.
 \end{equation}
Therefore, the normalization constant \(C_{\rm aft}\) and hence \(\lambda_{\rm aft}\) can be obtained from the measured \(P_{\rm aft}\) value.

The above definition of the parameter \(x\) does not account for seeds due to correlated noise, but their charge contribution is included in \(y\). Consequently, correlated noise may make \(y\) greater than \(x\) if nonlinearity is low.

\section{Simulation Results}\label{sec:simulation}

As mentioned above, in this work we used the MC code developed by \cite{jha2013simulating}, with some changes. First, we implemented rectangular, triangular, and exponential light pulses, in addition to the already available double exponential pulses, to study the effect of pulse shape on the nonlinear response of SiPMs. However, most of our results were obtained assuming simple exponential pulses. Second, we included an option to activate/deactivate the details of the SiPM electrical model described in \cite{jha2013simulating} (e.g., rise and decay times of charge avalanches, parasitic capacitance of several electronic elements, and resistance of the readout circuit). Unless otherwise stated, we deactivated these details and thus assumed the simplified model given by (\ref{overvoltage recovery}) and (\ref{I_q}), which sufficiently describes the main features of pixel recovery. Third, the probability that a seed triggers an avalanche, including seeds due to correlated noise, was set to \(\varepsilon(U)/\varepsilon(U_{\rm op})\), where the overvoltage dependence of the PDE was assumed to be given by (\ref{PDE}). Finally, we implemented delayed crosstalk from zero and parameterized the three components of correlated noise as described in section \ref{sec:overview}.

\begin{figure}[h]
\centering
\includegraphics[width=\linewidth]{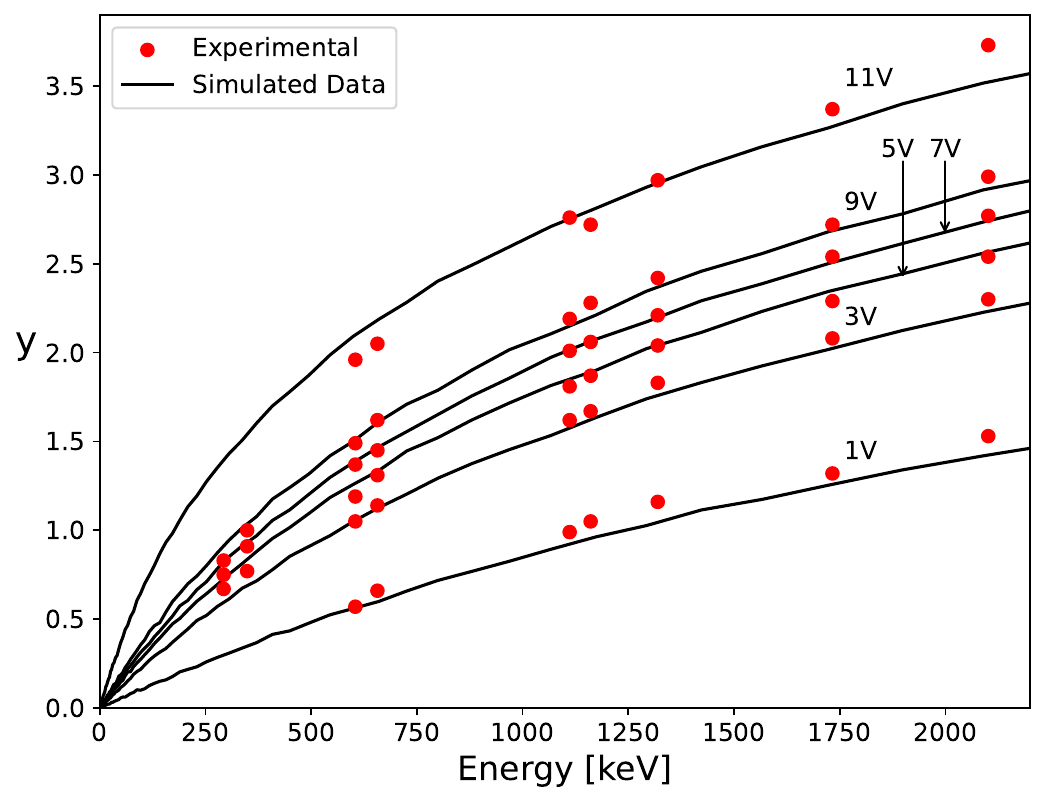}
\caption{Comparison of simulation results with experimental data for a Hamamatsu SiPM S13369-1350CS, taken from \cite{rosado2019modeling}. The output signal is expressed in terms of \(y=\langle Q\rangle/(N\,G\,e)\). See text for details.}
\label{fig:validation}
\end{figure}

To demonstrate the capability of this MC code to reproduce experimental data, we simulated a simple experiment described in \cite{rosado2019modeling}. This experiment involved a Hamamatsu SiPM S13369-1350CS optically coupled to a LYSO scintillator irradiated by gamma rays of different energies from \(300\) to \(2100\)~keV. The parameter \(y=\langle Q\rangle/(N\,G\,e)\) was measured as a function of gamma-ray energy at several \(U_{\rm op}\) values from \(1\)~V to \(11\)~V, as shown in Fig. \ref{fig:validation}. For the simulations, we set the following SiPM parameters:

\(N=667\), \(\tau_{\rm rec}=29\)~ns, \(\varepsilon_{\rm max}=0.597\), \(U_{\rm ch}=2.68\)~V, \(U_{0}=0\)~V, and \(\tau_{\rm bulk}=8.5\)~ns, which were taken from the SiPM datasheet and from \cite{rosado2015characterization}. For simplicity, delayed crosstalk was neglected, and \(P_{\rm aft}\) was assumed to be equal to \(P_{\rm ct}\), which is lower than \(5\%\) at \(U_{\rm op}\) values up to 7~V for this SiPM \cite{rosado2015characterization}. No direct measurements of \(P_{\rm ct}\) are available at higher \(U_{\rm op}\) values, because correlated noise increases dramatically, causing the large increase in \(y\) at \(11\)~V \cite{rosado2019modeling}. Therefore, \(P_{\rm ct}\) values at \(U_{\rm op}=9\) and \(11\)~V were adjusted so that simulations fit experimental data. We assumed exponential light pulses with a decay time of \(42\)~ns and a light yield of \(29\)~photons/keV for the LYSO crystal. However, the collection of photons on the SiPM was not well known, therefore it was fitted to experimental data, finding a value of \(13.8\%\) consistent with that obtained in \cite{rosado2019modeling}. For instance, for a gamma ray of \(662\)~keV, the mean number of photons incident on the SiPM is \(2650\). The number of simulated photons was sampled from a Poisson distribution, and they were randomly distributed among the \(667\) pixels of the SiPM. Results were obtained by averaging over a large number of simulations to smooth out statistical fluctuations.

The simulation results exhibit the correct nonlinear behavior and dependence on \(U_{\rm op}\). Remarkably, they reproduce the continuous increase in \(y\) with energy, in contrast to the generally assumed exponential saturation curve (\ref{instantaneous}). A more precise simulation of the SiPM response would require including the electrical details of the SiPM and the characteristics of the light pulses (e.g., rise time, wavelength spectrum, and light diffusion on the inner walls of the scintillator). However, the only aim of this example is to show that the simulation accounts for all the necessary elements to describe the nonlinear response of SiPMs. The key parameters in nonlinearity are analyzed below.

\subsection{Photon Rate}\label{ssec:photon_rate}

Nonlinearity depends on the balance between the rate of photons and the pixel recovery. If the photon rate is high enough so that photons may arrive at a pixel while it is recovering from a previous avalanche, no more avalanches are produced in this pixel or, if any is produced, it will have a lower charge multiplication. On the other hand, if the light pulse is very long, the average time elapsed between consecutive photons arriving at a same pixel is larger than the recovery time and nonlinearity will be weak, even though the mean number of photons per pixel is large. Actually, the relevant parameter is the mean number of seeds per pixel \(x = \varepsilon(U_{\rm op})\langle n\rangle/N\) rather than the mean number of photons per pulse.

\begin{figure}[h]
\centering
\includegraphics[width=\linewidth]{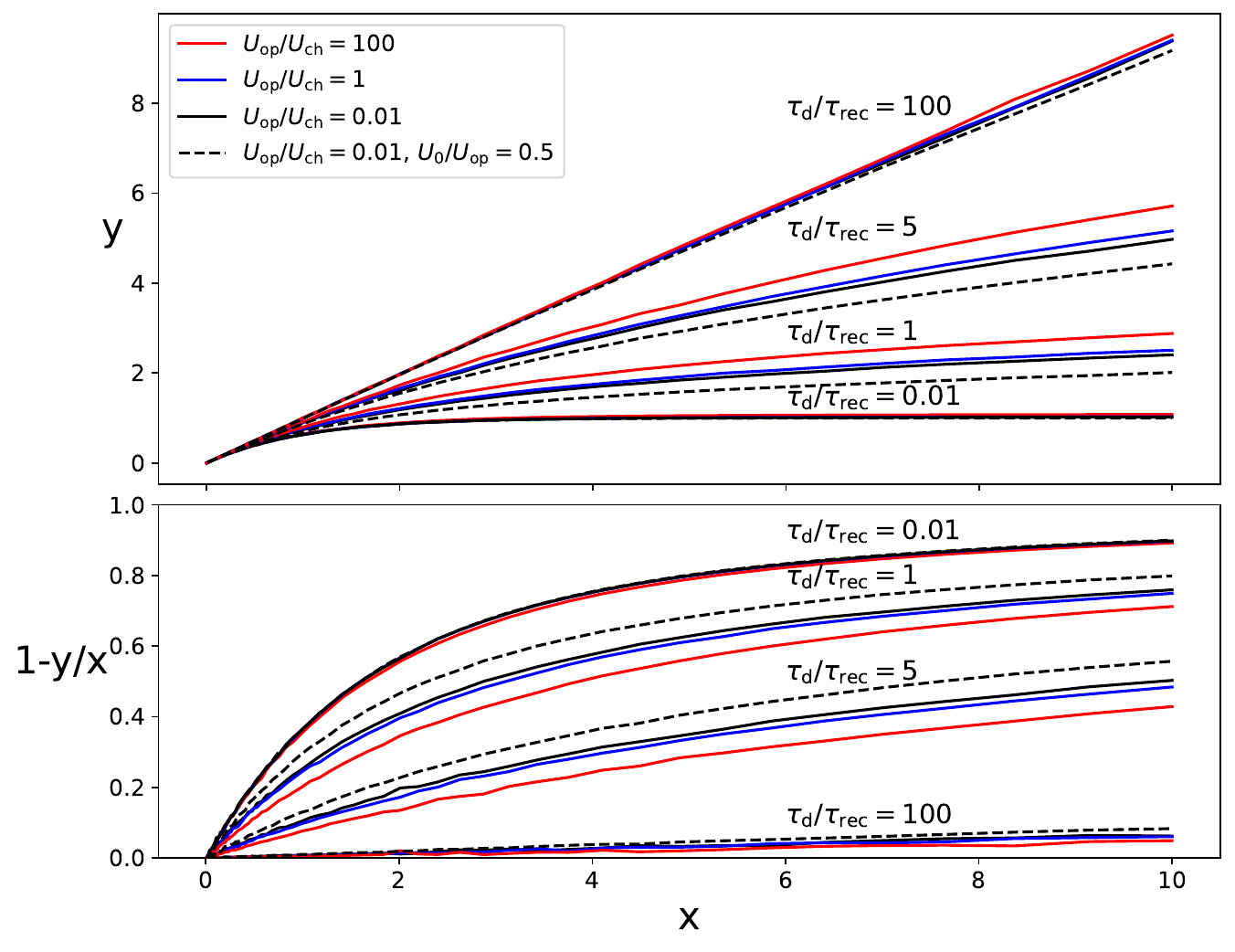}
\caption{Simulation results of the SiPM response for different values of \(\tau_{\rm d}/\tau_{\rm rec}\) and \(U_{\rm op}/U_{\rm ch}\) for exponential light pulses. To better understand nonlinearity, the output signal (upper plot) is expressed in terms of \(y=\langle Q\rangle/(N\,G\,e)\) and the incident light intensity in terms of the mean number of seeds per pixel \(x = \varepsilon(U_{\rm op})\langle n\rangle/N\). In the lower plot, the nonlinearity parameter \(1-y/x\) is shown.}
\label{fig:Td/Tr}
\end{figure}

We simulated exponential light pulses with different decay times \(\tau_{\rm d}\) incident in a SiPM with a fixed recovery time \(\tau_{\rm rec}\). For these tests, we used the same simulation parameters as in Fig. \ref{fig:validation}, but ignoring correlated noise and assuming \(U_{\rm op}\gg U_{\rm ch}\), that is, the PDE reaches the maximum value \(\varepsilon_{\rm max}\). Results of \(y\) versus \(x\) are shown as red lines in the upper plot of Fig. \ref{fig:Td/Tr}. In the lower plot, it is shown the nonlinearity, defined as \(1-y/x\). For \(\tau_{\rm d}/\tau_{\rm rec}=100\), nonlinearity is smaller than \(5\%\) up to \(x=11\). The smaller \(\tau_{\rm d}/\tau_{\rm rec}\), the smaller the \(x\) value at which such nonlinearity of \(5\%\) is reached, \(y\) showing a logarithmic-like behavior with \(x\). In the limit \(\tau_{\rm d}\ll\tau_{\rm rec}\), results tend to (\ref{instantaneous}), where nonlinearity reaches \(5\%\) at an \(x\) value as small as \(0.1\). Results were checked to not depend on the particular values of \(\tau_{\rm d}\) and \(\tau_{\rm rec}\) for a fixed \(\tau_{\rm d}/\tau_{\rm rec}\) ratio.

\subsection{Operation overvoltage}\label{ssec:overvoltage}

As explained above, the probability that a seed triggers a charge avalanche in a recovering pixel is assumed to be \(\varepsilon(U)/\varepsilon(U_{\rm op})\), where \(U\) recovers to \(U_{\rm op}\) with the time elapsed from the last avalanche as given by (\ref{overvoltage recovery}) and (\ref{I_q}). In the limit \(U_{\rm op}\gg U_{\rm ch}\), this trigger probability tends to a Heaviside step function of time, that is, it recovers to unity almost instantaneously after each avalanche. Therefore, nonlinearity is only due to the uncompleted recovery of the charge multiplication in subsequent avalanches (\ref{gain_recovery}). In the opposite limit  \(U_{\rm op}\ll U_{\rm ch}\) (assuming \(U_0=0\)~V), both the trigger probability and the charge multiplication are proportional to \(U\). In this situation, nonlinearity would be stronger for the same number of seeds, although operating a SiPM at such low overvoltage is unusual, because the PDE is too low.

We performed the simulations described in section \ref{ssec:photon_rate} for different \(U_{\rm op}/U_{\rm ch}\) ratios. Results for \(U_{\rm op}/U_{\rm ch}=0.01,\,1\) and \(100\) are shown as solid lines in Fig. \ref{fig:Td/Tr}. The impact of \(U_{\rm op}/U_{\rm ch}\) on the nonlinearity is not so high as that of \(\tau_{\rm d}/\tau_{\rm rec}\), but it cannot be ignored. Again, results were checked to not depend on \(U_{\rm op}\) or \(U_{\rm ch}\) for a fixed \(U_{\rm op}/U_{\rm ch}\) ratio. Remarkably, it can be seen that the effect of increasing \(U_{\rm op}/U_{\rm ch}\) is almost equivalent to a small increase in \(\tau_{\rm d}/\tau_{\rm rec}\), proving that nonlinear effects are essentially due to the balance between the photon rate and the pixel recovery, regardless of how the recovery takes places, i.e., only for charge multiplication or for both charge multiplication and trigger probability.

The threshold overvoltage \(U_0\) is usually very small compared to \(U_{\rm op}\) and can be generally ignored. In case that \(U_0\) is significant, the trigger probability is zero in the first stage of the pixel recovery while \(U\leq U_0\), causing that nonlinearity to be stronger. Results for \(U_{\rm op}/U_{\rm ch}=0.01\) and \(U_0/U_{\rm op}=0.5\) are shown as dashed lines in  Fig. \ref{fig:Td/Tr} to illustrate this effect.

\subsection{Correlated noise}\label{ssec:correlated_noise}

Correlated noise can be understood as a process causing the mean number of seeds per pixel to increase for the same number of incident photons. As a consequence, its main effect is that the effective gain of the SiPM in the linear region is greater than the nominal gain \(G\), meaning that \(y\) is increased by an extra gain factor that we refer to as \(1+c\). Both crosstalk components have a larger contribution to this gain factor than afterpulsing for the same probability, because the mean multiplication charge of avalanches due to afterpulsing is smaller, as they are usually produced while the pixel is still recovering. For exponential pulses with \(\tau_{\rm d}=5\)~ns incident on a SiPM with \(N=100\), \(\tau_{\rm rec}=\tau_{\rm bulk}=10\)~ns and \(U_0=0\), and taking \(U/U_{\rm ch}=1 \) and \(x\ll 1\), we found that a crosstalk probability of \(10\%\) (irrespective of whether it is prompt or delayed crosstalk) leads to \(c=0.101\), whereas \(c=0.059\) is obtained for the same probability of afterpulsing. Notice that \(c\) can be slightly greater than the probability of crosstalk, which is defined as the probability of inducing one or more secondary avalanches. This is partly compensated by the fact that pixels at the borders of the SiPM have a lower crosstalk probability. The higher the number of pixels, the smaller these border effects. On the other hand, cascades of secondary avalanches due to any component of correlated noise contribute to increasing \(c\), this effect being relevant when the probability of correlated noise is high, as discussed in \cite{rosado2019modeling}. Analytical expressions of \(c\) for crosstalk are available in \cite{gallego2013modeling, vinogradov2012analytical}, but the accurate value of \(c\) should be obtained by simulation when afterpulsing is not negligible.

\begin{figure}[t]
\centering
\includegraphics[width=\linewidth]{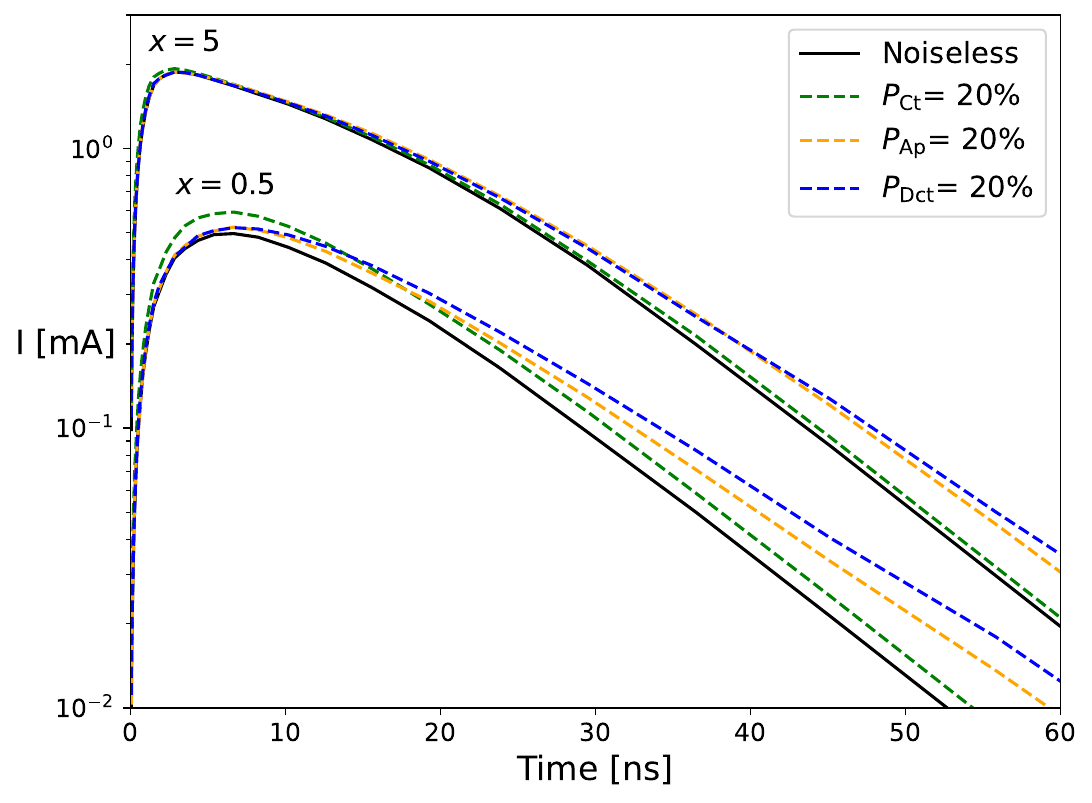}
\caption{Averaged SiPM signals simulated for different situations of correlated noise (i.e., no noise, \(20\%\) of prompt crosstalk, \(20\%\) of afterpulsing, or \(20\%\) of delayed crosstalk) and different amounts of seeds per pixel (i.e., \(x=0.5\) and \(5\)).}
\label{fig:signals}
\end{figure}

At high photon rates, where nonlinearity is significant, the uncompleted recovery of pixels reduces the gain factor due to correlated noise. For exponential light pulses, prompt crosstalk was found to be more mitigated than delayed crosstalk and afterpulsing for the same probability. This is because seeds due to prompt crosstalk are mainly produced at the beginning of the light pulse, just when the photon rate is highest. In contrast, seeds due to delayed crosstalk and afterpulsing are produced later, when the photon rate is lower and the recovery of pixels is more advanced. These effects are more noticeable when studying the shape of the output signal rather than the integrated charge. In Fig. \ref{fig:signals}, averaged simulated signals for a probability of correlated noise of \(20\%\) (assuming a sole noise component in each case) are compared with that obtained in the absence of correlated noise. We used the same simulation parameters as before, i.e., \(\tau_{\rm d}=5\)~ns and \(\tau_{\rm rec}=\tau_{\rm bulk}=10\)~ns. Prompt crosstalk gives rise to an almost constant scale factor to the signal at \(x=0.5\), but the effect is much lower at \(x=5\) as a consequence of nonlinearity. Both afterpulsing and delayed crosstalk produce a lengthening of the output current pulse, which can be significant if \(\tau_{\rm bulk}\) is large. This lengthening, along with the distortion due to nonlinearity, makes the signal pulse to be flatter than the incident light pulse.

\subsection{Pulse shape}\label{ssec:pulse_shape}

We also studied the influence of the shape of the incident light pulse on nonlinearity by comparing the SiPM response for exponential, rectangular, triangular, and double exponential pulses. For a direct comparison, pulses should have the same mean photon rate for a given number \(n\) of photons per pulse. The probability density distribution (PDF) and the mean photon rate for the four pulse shapes are shown in Table \ref{tab:pulse_shapes}.

\begin{table}
    \centering\small
    \renewcommand{\arraystretch}{1.5}
    \begin{tabular}{|c|c|c|} \hline 
         \textbf{Shape} & \textbf{Probability density function} & \textbf{Mean ph. rate} \\ \hline 
         Exponential & $\frac{1}{\tau_{\rm d}}e^{-t/\tau_{\rm d}} \;\; (t \geq 0)$ & $\frac{n}{2\tau_{\rm d}}$ \\ \hline 
         Rectangular & $\frac{1}{\tau_{\rm w}} \;\; (0 \leq t \leq \tau_{\rm w})$ & $\frac{n}{\tau_{\rm w}}$ \\ \hline 
         Double exp. & $\frac{1}{\tau_2-\tau_1}\left(e^{-t/\tau_2}-e^{-t/\tau_1}\right) \;\; (t \geq 0)$ & $\frac{n}{2(\tau_1+\tau_2)}$ \\ \hline
         \multirow{2}{*}{Triangular} & $\frac{2}{\tau_{\rm r}+\tau_{\rm f}}\frac{t}{\tau_{\rm r}} \;\; (0 \leq t \leq \tau_{\rm r})$ & \multirow{2}{*}{$\frac{4n}{3(\tau_{\rm r}+\tau_{\rm f})}$} \\ 
          & $\frac{2}{\tau_{\rm r}+\tau_{\rm f}}\left(1-\frac{t}{\tau_{\rm f}}\right) \;\; (\tau_{\rm r} < t \leq \tau_{\rm r}+\tau_{\rm f})$ & \\ \hline
    \end{tabular}
    \caption{Characteristics of different pulse shapes.}
    \label{tab:pulse_shapes}
\end{table}

\begin{figure}[t]
\centering
\includegraphics[width=\linewidth]{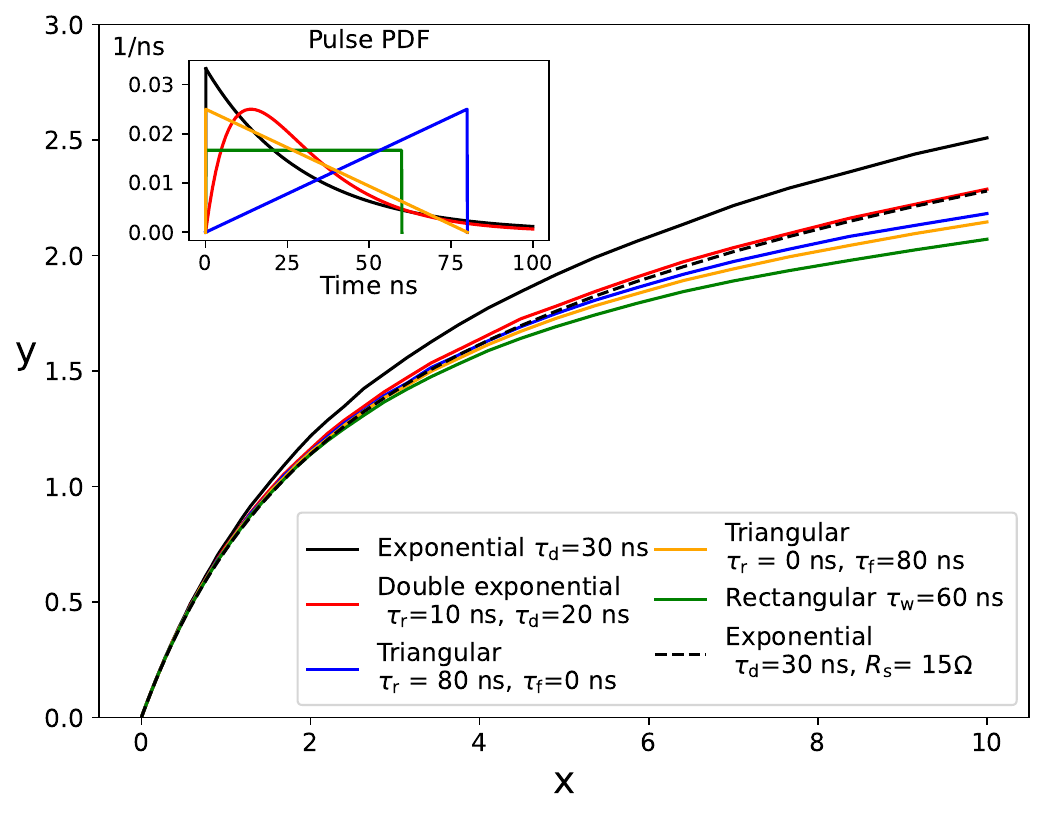}
\caption{Simulated SiPM response for light pulses of different shapes but same mean photon rate. The inset plot shows the probability density functions of the light pulses used in the simulations. The dashed line represents the results for an exponential pulse including the detailed electrical model developed in \cite{jha2013simulating}, instead of the simplified one described in section \ref{sec:overview}.}
\label{fig:various_effects}
\end{figure}

Simulation results for different pulses with mean photon rate \(n/\tau\), taking \(\tau=30\)~ns, are shown as solid lines in Fig. \ref{fig:various_effects}. We included exponential pulses of decay time \(\tau_{\rm d}=30\)~ns, rectangular pulses of width \(\tau_{\rm w}=60\)~ns, rising triangular pulses with \(\tau_{\rm r}=80\)~ns and \(\tau_{\rm f}=0\), falling triangular pulses with \(\tau_{\rm r}=0\) and \(\tau_{\rm f}=80\)~ns, and double exponential pulses with \(\tau_1=10\)~ns and \(\tau_2=20\)~ns. The PDF for these pulses is shown in the inset of the figure for a better comparison. For these simulations, we assumed \(N=100\), \(\tau_{\rm rec}=30\)~ns, \(U_{\rm op}/U_{\rm ch}=1\), \(U_0=0\)~V and neglected correlated noise. Results are similar for all the pulse shapes, but nonlinearity is stronger for triangular and rectangular pulses, which have a finite duration.

\begin{figure*}[t]
\centering
\includegraphics[width=\linewidth]{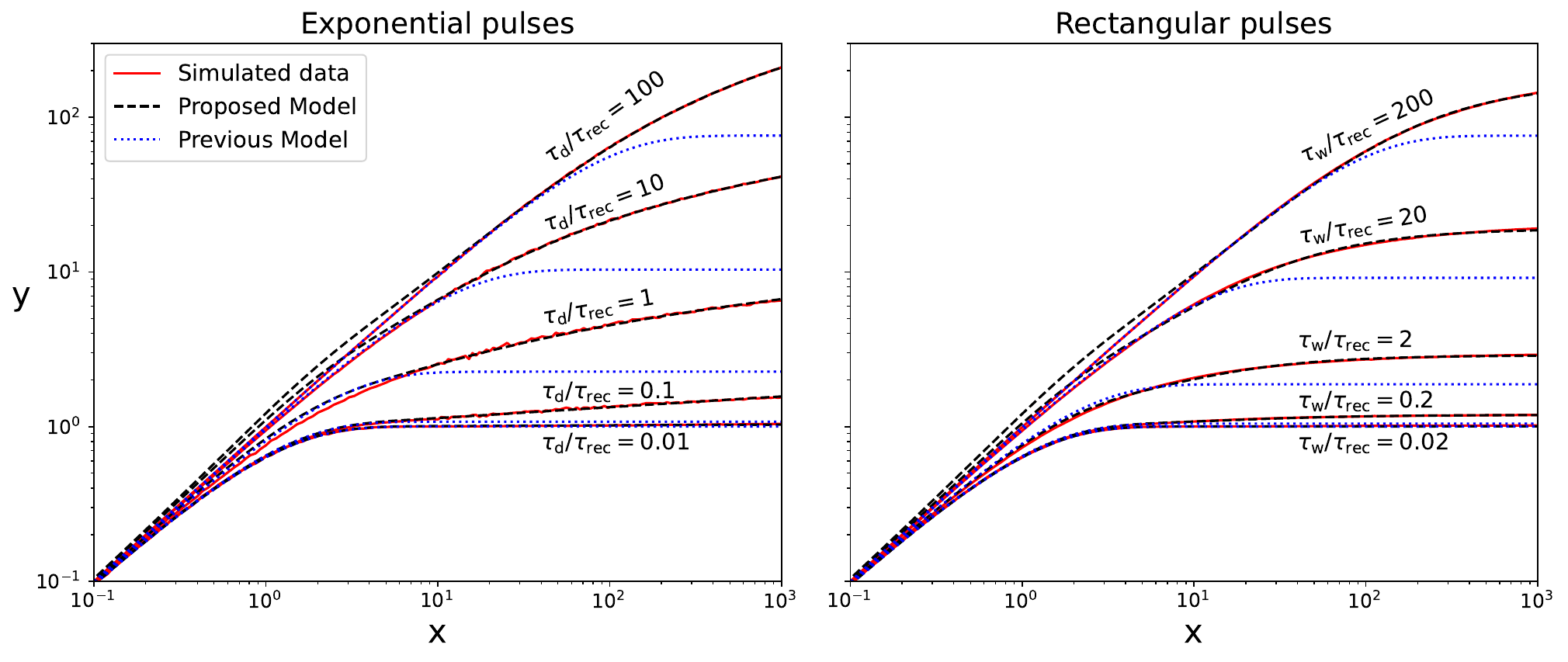}
\caption{Simulation data for both exponential pulses with different \(\tau_{\rm d}/\tau_{\rm rec}\) ratios (left-hand plot) and rectangular pulses with different \(\tau_{\rm w}/\tau_{\rm rec}\) ratios (right-hand plot). Fits of the models (\ref{model_2019}) and either (\ref{model_exp}) or (\ref{model_finite}), depending on the case, are also shown. Results are depicted for a large range of \(x\), but the model (\ref{model_2019}) was fitted in a limited range where nonlinearity is still moderate. The models (\ref{model_exp}) and (\ref{model_finite}) outperform the model (\ref{model_2019}) for both types of pulses.}
\label{fig:fits}
\end{figure*}

Further results for a larger \(x\) range from \(0\) to \(1000\) and different \(\tau/\tau_{\rm rec}\) ratios are shown as red solid lines in Fig. \ref{fig:fits}. Results for both pure exponential pulses with \(\tau_{\rm d}=\tau/2\) and rectangular pulses with \(\tau_{\rm w}=\tau\) are shown in the left-hand plot and the right-hand plot, respectively. Again, we assumed \(N=100\), \(U_{\rm op}/U_{\rm ch}=1\), \(U_0=0\)~V and neglected correlated noise. Notice that the SiPM response saturates at a certain level when increasing \(x\) for rectangular pulses. We checked that the same behavior is shown for triangular pulses. On the other hand, the response is found to have a logarithmic growth with \(x\) for exponential-like pulses (including double exponential ones), except in the limit \(\tau\ll\tau_{\rm rec}\). This is due to the fact that, for exponential-like pulses, a few photons may arrive at long times producing charge avalanches in fairly recovered pixels.

\subsection{Electrical model}\label{ssec:electrical_model}

For all the simulation cases shown so far, we assumed the simplified electrical model described in section \ref{sec:overview}. Under these simplifications, results expressed in terms of the parameters \(x\) and \(y\) are independent of the number of pixels \(N\), except for the above-mentioned small border effects on crosstalk. However, when \(N\) is increased, the sum output current is higher for the same \(x\) value and gain. As a consequence, the voltage drop across the readout-circuit resistance \(R_{\rm s}\) cannot be longer ignored and the overvoltage at the pixels is reduced, as discussed in \cite{jha2013simulating}. The parasitic capacitance due to the unfired pixels connected in parallel to the fired pixels may also be relevant for large \(N\), increasing the recovery time and hence contributing to nonlinearity as well.

We performed simulations for exponential pulses including the detailed electrical model developed by \cite{jha2013simulating}. We assumed a readout-circuit resistance \(R_{\rm s}=15\,\Omega\) and typical values of the parameters used in this electrical model. For \(N=100\), results were almost identical to those obtained with the simplified electrical model (black solid line in Fig. \ref{fig:various_effects}), but nonlinearity is found to be stronger for \(N=3600\) (black dashed line), as expected. The resulting curve happens to be very similar to that obtained for double exponential pulses with the simplified electrical model (red line), suggesting that the SiPM response may be described by a universal function characterized by a few parameters.

\section{Analytical Model}\label{sec:model}

In \cite{rosado2019modeling}, a statistical model including the main factors affecting the nonlinearity (i.e., photon rate, overvoltage, correlated noise, and pulse shape) was developed. From this model, it was derived the following simple expression for the response of SiPMs at moderate nonlinearity:
\begin{equation}\label{model_2019}
    y=\frac{1}{1-\gamma}\,\left\{1-\exp\left[-(1-\gamma)(1+c)\,x\right]\right\}\,,
\end{equation}
where \(1+c\) is the gain factor due to correlated noise explained in section \ref{ssec:correlated_noise}, and the dimensionless parameter \(\gamma\) is given by
\begin{equation}\label{gamma}   \gamma=2\int_{0}^{\infty}\int_{0}^{\infty}p(t)\,p(t+t_{\rm s})\,\frac{\varepsilon(U(t_{\rm s}))}{\varepsilon(U_{\rm op})}\,\frac{U(t_{\rm s})}{U_{\rm op}}\,{\rm d}t_{\rm s}\,{\rm d}t
\end{equation}
Here, \(t\) is the arrival time of a first seed in a pixel, \(t_{\rm s}\) is the time of a second seed in the same pixel with respect to the first one, and \(p(t)\) stands for the PDF of the output signal pulse, including the lengthening caused by afterpulsing and delayed crosstalk. The parameter \(\gamma\) represents the average charge in units of \(G\,e\) due to a seed produced in a recovering pixel, which ranges from \(0\) for very short pulses to \(1\) for very long pulses for which nonlinearity is negligible. In practice, both \(c\) and \(\gamma\) can be obtained by fitting (\ref{model_2019}) to experimental data instead of calculating them from more basic parameters that may be not well known, also taking into account that the model is based on several approximations.

In figure \ref{fig:fits}, fits of the model (\ref{model_2019}) to simulated data for pure exponential pulses and rectangular pulses are shown as blue dotted lines in the left-hand plot and right-hand plot, respectively. Although results are shown up to \(x=1000\), these fits were done in a limited range of \(x\), depending on \(\tau/\tau_{\rm rec}\), where nonlinearity is weak. Notice that \(\gamma\) accounts for interactions between only two seeds in a same pixel, hence this model is expected to be valid only for \(x\lesssim 2\). Indeed, it predicts an exponential saturation in the limit \(x\rightarrow\infty\) that is actually not observed, especially for exponential-like pulses.

We extended the above model to arbitrary \(x\) values based on simulation results. In the first place, \(y\rightarrow(1+c)\,x\) should be fulfilled in the limit \(x\rightarrow 0\) as well as for very long pulses, regardless of the pulse shape. In the second place, \(y\) saturates as \(1-\exp(-x)\) for very short pulses. In the third place, \(y\) was found to have a logarithmic growth with \(x\) for exponential-like pulses, whereas it saturates at a certain value depending on \(\tau/\tau_{\rm rec}\) for pulses of finite width \(\tau\), but more slowly than the model (\ref{model_2019}) predicts. Given these conditions, we propose the following ansatz for exponential-like pulses:
\begin{equation} \label{model_exp}
y=\left[1+c+a_{\rm e}\,\ln{\left(1+b_{\rm e}\, x \right)}\right] \left[1-\exp{\left(-x\right)}\right]\,,
\end{equation}
and this one for finite pulses:
\begin{equation} \label{model_finite}
y=\left[1+c+\frac{a_{\rm f}\,x}{1+b_{\rm f}\, x}\right] \left[1-\exp{\left(-x\right)}\right]\,,
\end{equation}
where we used the indexes ``e'' and ``f´´ to distinguish between the two categories of pulses.

Fits of these new two models are shown as black dashed lines in Fig. \ref{fig:fits}. Both models fit adequately to simulated data in the entire \(x\) range from \(0\) to \(1000\), with residuals being generally less than \(5\%\), although depending on both the ratio of the pulse width to \(\tau_{\rm rec}\) and the selected fitting range. For instance, the maximum residual for exponential pulses with \(\tau_{\rm d}/\tau_{\rm rec}=1\) is \(10\%\) when fitting (\ref{model_exp}) in the entire range, but the maximum residual is as low as \(3.4\%\) when the fit is limited to \(x\) values smaller than \(10\), where nonlinearity reaches \(70\%\) (see Fig. \ref{fig:Td/Tr}). The fit is poorer for \(\tau_{\rm d}/\tau_{\rm rec}=100\), where nonlinearity is only noticeable at very large \(x\) values. Similar residuals are obtained for double exponential pulses as well as when fitting the model (\ref{model_finite}) for finite pulses. We also checked that the models (\ref{model_exp}) and (\ref{model_finite}) fit well to simulated data for all the variety of simulation cases studied in this work (e.g., different \(U_{\rm op}/U_{\rm ch}\) values and probabilities of correlated noise).

Although the models (\ref{model_exp}) and (\ref{model_finite}) were obtained in a heuristic way, we can give a meaning to their different terms. The factor \(1-\exp(-x)\) in both models is the mean fraction of pixels with one or more seeds produced by photons, which is equal to the mean fraction of fired pixels per pulse in the absence of crosstalk. On the other hand, the factor \(1+c+a_{\rm e}\ln{\left(1+b_{\rm e}\, x \right)}\) for exponential-like pulses and the factor \(1+c+\frac{a_{\rm f}\,x}{1+b_{\rm f}\, x}\) for finite pulses correspond to the mean charge in units of \(G\,e\) per fired pixel, which should be always greater than unity, because at least the first avalanche in any fired pixel has a charge equal to \(G\,e\). The parameter \(c\) is the contribution of correlated noise induced by this first avalanche in both models, while the term either \(a_{\rm e}\ln{\left(1+b_{\rm e}\, x \right)}\) or \(\frac{a_{\rm f}\,x}{1+b_{\rm f}\, x}\) accounts for all the remaining avalanches produced during the pixel recovery, therefore having a similar meaning to (\ref{gamma}). Indeed, the relationships \(a_{\rm e}b_{\rm e}\approx a_{\rm f}\approx \frac{\gamma}{2}\) are found by comparing the Taylor series of (\ref{model_2019}), (\ref{model_exp}) and (\ref{model_finite}) at \(x=0\) in the absence of correlated noise. We checked that the fitting parameters of the three models depend on \(\tau/\tau_{\rm rec}\) and \(U_{\rm op}/U_{\rm ch}\) in a consistent way when the fit is limited to a small \(x\) range, where the models are equivalent.

\begin{figure}[h]
\centering
\includegraphics[width=\linewidth]{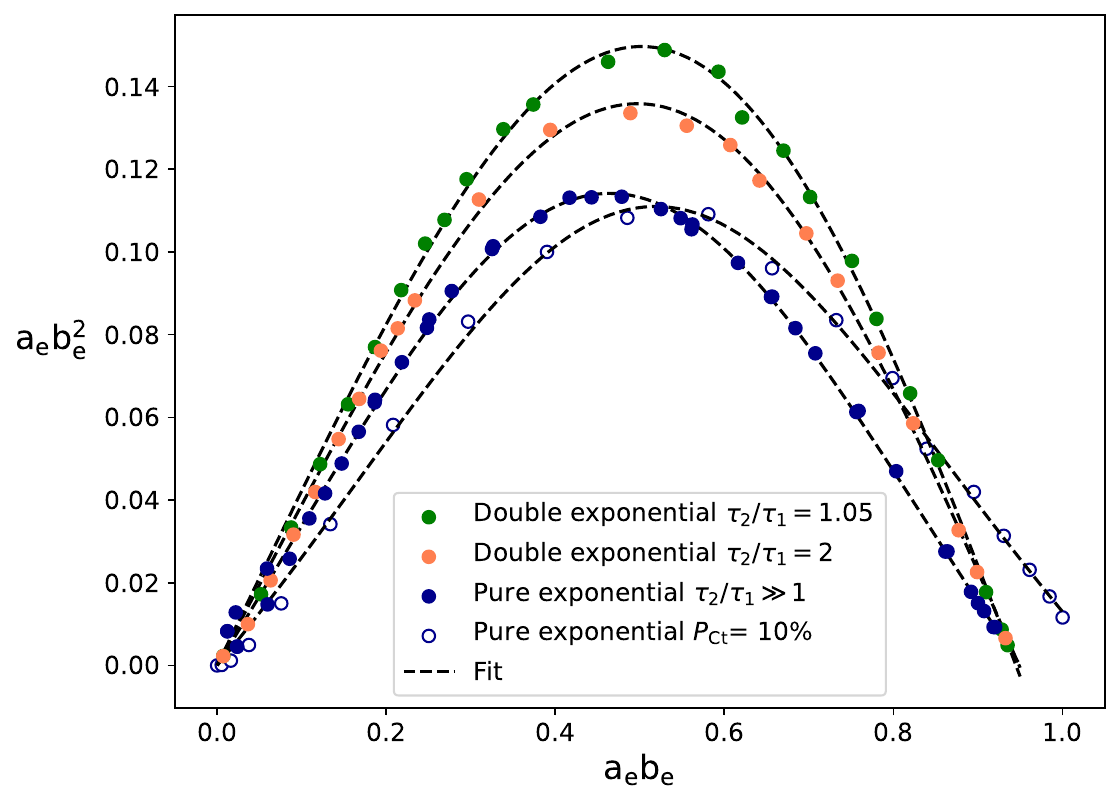}
\caption{Parameters of the model (\ref{model_exp}) fitted to simulation data for exponential-like pulses with different shapes as well as for different combinations of \(t_{\rm rec}\) and \(U_{\rm op}\). Filled circles and empty circles correspond to simulated data without correlated noise and with a probability of prompt crosstalk of \(10\%\), respectively. The relationship between these parameters is correctly described by (\ref{ab_fit}) for a given pulse shape and probability of correlated noise.}
\label{fig:ab}
\end{figure}

When considering a large \(x\) range, the various dependencies of the SiPM response are contained in the pair of fitting parameters either \((a_{\rm e},\,b_{\rm e})\) or \((a_{\rm f},\,b_{\rm f})\) in a more complex way. This is illustrated in Fig. \ref{fig:ab}, where the parameters \((a_{\rm e},\,b_{\rm e})\) obtained by fitting (\ref{model_exp}) to simulation data for exponential-like pulses with different shapes (i.e., several \(\tau_2/\tau_1\) ratios for double exponential pulses) and different combinations of \(\tau_{\rm rec}\) and \(U_{\rm op}\) are shown. In most cases, no correlated noise was included, but we also simulated cases with a probability of prompt crosstalk of \(10\%\). The fitting parameters vary somewhat depending on the considered \(x\) range. For the data shown in the figure, all the fits were done in the range \(0\leq x\leq100\). Instead of plotting \(b_{\rm e}\) versus \(a_{\rm e}\), we plotted \(a_{\rm e}b_{\rm e}^2\) versus \(a_{\rm e}b_{\rm e}\), which are the coefficients of the first two terms of the Taylor series of \(a_{\rm e}\ln{\left(1+b_{\rm e}\, x \right)}\). The limit \(a_{\rm e}b_{\rm e}\rightarrow 0\) corresponds to very short pulses, while \(a_{\rm e}b_{\rm e}\) is close to unity for very long pulses. We found that the relationship between \(a_{\rm e}b_{\rm e}\) and \(a_{\rm e}b_{\rm e}^2\) can be properly described by:
\begin{equation}\label{ab_fit}
    a_{\rm e}b_{\rm e}^2=\frac{k_1 (a_{\rm e}b_{\rm e})-k_2 (a_{\rm e}b_{\rm e})^2}{1-k_3 (a_{\rm e}b_{\rm e}) + k_4 (a_{\rm e}b_{\rm e})^2}\,,
\end{equation}
where the parameters \(k_1,\,k_2,\,k_3\) and \(k_4\) depend on the particular pulse shape and the probability of correlated noise.

Although this relationship is not straightforward, it proves the consistency of the model, which provides a family of curves for the SiPM response in a large variety of situations. A very similar result is obtained for finite pulses when plotting \(a_{\rm f}b_{\rm f}\) as a function of \(a_{\rm f}\).

\section{Conclusions}\label{sec:conclusions}

In this work, we carried out a systematic study of the nonlinear response of SiPMs for light pulses, identifying and parameterizing the main factors affecting nonlinearity. For this purpose, we used the MC code developed by \cite{jha2013simulating}, with some improvements in the description of the correlated noise. Simulations were shown to reproduce experimental data on the output charge for scintillation light pulses as a function of both the pulse intensity and the SiPM operation overvoltage.

We observed that the SiPM response is quite universal when the output charge is expressed as the equivalent number of full charge avalanches per pixel (dubbed \(y\)) as a function of the expected number of avalanches per pixel for an ideal SiPM in the absence of nonlinearity and correlated noise (dubbed \(x\)).

Nonlinearity essentially depends on the balance between the photon rate and the pixel recovery time \(\tau_{\rm rec}\) after a charge avalanche, which can be parameterized as the ratio of the pulse width \(\tau\) (e.g., the decay time for an exponential pulse) to \(\tau_{\rm rec}\) for a given \(x\) value. For \(\tau/\tau_{\rm rec}\approx1\), nonlinearity reaches around \(5\%\) at \(x\) values as low as \(0.2\).

When a SiPM is operated at a high overvoltage, such that the PDE is close to its maximum value, the trigger probability of a pixel recovers faster than the pixel overvoltage, reducing nonlinearity in a similar way as if \(\tau_{\rm rec}\) is made shorter. We parameterized it as the ratio of the operation overvoltage to the characteristic parameter \(U_{\rm ch}\) of the PDE curve assuming (\ref{PDE}), albeit other parameters should be used instead in case that the PDE is described by a different function.

We also studied the SiPM response for different light pulse shapes, finding that nonlinearity is slightly stronger for finite pulses (e.g., rectangular and triangular ones) than exponential-like pulses with the same mean photon rate. In addition, we demonstrated that the SiPM response saturates at a certain level when increasing the light pulse intensity for finite pulses, whereas it shows a logarithmic growth for exponential-like ones.

The correlated noise was found to basically increase the effective gain of the SiPM in the linear region, having a minor influence on nonlinearity. However, afterpulsing and delayed crosstalk may still be relevant for intense light pulses, because these noise components lead to a lengthening of the output signal. We showed how the signal is distorted by nonlinearity and correlated noise. The impact of the details of the electrical model on simulation results has also been studied, confirming the findings shown in \cite{jha2013simulating}.

Based on these simulation results and the statistical model developed by \cite{rosado2019modeling}, we found the two simple expressions (\ref{model_exp}) and (\ref{model_finite}) that reproduce correctly all the features of the SiPM response for exponential-like and finite light pulses, respectively. Even though these models have only three fitting parameters, they were shown to be valid even for light pulses of very high intensity and for a wide range of situations. The different terms of both models were properly explained and the relationship between the fitting parameters was assessed and understood.

\section*{Acknowledgment}

This work was supported by the Spanish Research State Agency (AEI) through the grant PID2019-104114RB-C32. V.~Moya also appreciates a doctoral fellowship provided by IPARCOS Institute and the research grant CT19/23-INVM-109 funded by NextGenerationEU. We thank Abhinav K.~Jha for providing help and guidance for the use of his MC code.

\bibliographystyle{IEEEtran}
\bibliography{arXiv.bbl} 
\end{document}